\def\simless{\mathbin{\lower 3pt\hbox
     {$\rlap{\raise 5pt\hbox{$\char'074$}}\mathchar"7218$}}}   
\def\simmore{\mathbin{\lower 3pt\hbox
     {$\rlap{\raise 5pt\hbox{$\char'076$}}\mathchar"7218$}}}   

\documentclass{ws-p9-75x6-50}
\def\Msun{{\rm M}_{\odot}}                                     
\def\hide#1{}

\begin{document}


\title{Kilohertz quasi-periodic oscillations and strong field gravity
in X-ray binaries}

\author{Mariano M\'{e}ndez}

\address{Astronomical Institute, University of Amsterdam, the
Netherlands 
and
SRON, National Institute for Space Research, 
Utrecht, the Netherlands. E-mail:
M.Mendez@SRON.nl}


\maketitle

\abstracts{In the past five years observations with the {\em Rossi
X-ray Timing Explorer} have revealed fast quasi-periodic oscillations
in the X-ray flux of about 20 X-ray binaries. Thought to originate
close to the surface of a neutron star, these oscillations provide
unique information about the strong gravitational field in which they
are produced.}


The dynamical timescale close to an object of mass $M$ and radius $R$
is $\tau \sim R / v_{\rm ff} = \sqrt{ R^3 / 2 G M}$, where $v_{\rm ff}$
is the local free-fall velocity. For typical masses and radii of
neutron stars ($M \sim 1-2$\,$\Msun$ and $R \sim 10-20$\,km), $\tau
\simless 10^{-3}$\,s. It was only after NASA's {\em Rossi X-ray Timing
Explorer (RXTE)} was launched in December 1995 that we could finally
probe these short timescales, both in neutron-star and black-hole X-ray
binaries. Groups at Goddard Space Flight Center and the University of
Amsterdam, almost simultaneously announced the detection of
submillisecond quasi-periodic variability in the X-ray flux of two
binary systems with neutron star primary\cite{iauc6319,iauc6320}. Since
then, so-called kilohertz quasi-periodic oscillations (kHz QPOs) have
been observed in more than 20 X-ray binary systems\cite{vdk00}.


KHz QPOs usually appear in pairs\cite{ford97a}, with frequencies
$\nu_{1}$ and $\nu_{2} (> \nu_{1})$ between $\sim 300$
Hz\cite{jonker00} (the lowest value for $\nu_{1}$) and $\sim 1330$
Hz\cite{straaten00} (the highest value for $\nu_{2}$). For a given
source both $\nu_{1}$ and $\nu_{2}$ can drift by $\sim 100$ Hz on
timescales of a few hundred seconds\cite{strohmayer96}, with a typical
dynamic range of $\sim 200 - 300$ Hz; up to $\sim 800$ Hz variations
have been observed in one source\cite{straaten00}. Despite these rather
large $\nu_{1}, \nu_{2}$ shifts, for each source $\Delta \nu = \nu_{2}
- \nu_{1}$, remains more or less 
constant\cite{ford97a,strohmayer96,wijnands98a}. In the best studied
sources, $\Delta \nu$ has been observed to decrease by $50-100$ Hz as
$\nu_{1}, \nu_{2}$ increase\cite{vanderklis97,mendez98c}. In all the
other sources of kHz QPO, $\Delta \nu$ is consistent with this
trend\cite{psaltis98}

It is widely accepted that at least one of the two QPO peaks is
produced by matter in Keplerian orbits around the neutron star at the
inner edge of the accretion disc$^{12-14}$.
If this is so, QPO frequencies should depend sensitively upon the mass
and angular momentum of the compact star as well as upon the orbital
radius at which the QPOs are produced\cite{mlp98a,sv99}. In principle
it is possible to measure some of these parameters
independently$^{23-25}$,
which would allow us to set simultaneous constraints to the mass and
radius of the star, and would provide means of knowing the high-density
equation of state of neutronic matter\cite{mlp98a}, an issue of central
importance in nuclear physics research.

On timescales of a few hours QPO frequencies correlate with X-ray
intensity\cite{ford97a}, but this correlation breaks down on timescales
of a day or more\cite{zhang98a}; on longer timescales QPO frequencies
correlate much better with the X-ray spectral properties of the
source$^{6,10-16}$.
It is thought that mass accretion rate is the mechanism behind this
behavior\cite{mlp98a}.

The Fourier power spectra of X-ray binaries often show a broad-band
component that is roughly flat below a break frequency at $\nu_{\rm b}
\sim 1 - 10$ Hz and decreases above $\nu_{\rm b}$, and a low-frequency
QPO at $\sim 10 - 50$ Hz. Studies with RXTE have recently shown that
both $\nu_{\rm b}$ and the frequency of the low-frequency QPO are
strongly correlated to the frequency of the kHz QPOs$^{16,26-28}$
and to each other\cite{wk99}, and that these correlation encompasses
systems both with neutron star and black hole primaries. This result is
quite significant, as it seems to indicate that the presence of a solid
surface, a magnetic field, or an event horizon plays no significant
role in the mechanism that produces the rapid variability observed in
these objects.

\section*{Acknowledgments} This work was supported by the Netherlands
Organization for Scientific Research, grant PGS 78-277, the Netherlands
Foundation for research in astronomy, grant 781-76-017, the Netherlands 
Research School for Astronomy, and by LKBF. The author is grateful to 
Max-Planck-Institut f\"ur Astrophysik for their hospitality.

\end{document}